# A Novel Scheme of Digital Instantaneous Automatic Gain Control (DIAGC) for Pulse Radars

Sumanta Pal, Nirmala Shanmugam, Mohit Kumar, P Radhakrishna
Electronics and Radar Development Establishment
Defense Research and Development Organization
Bangalore – 560093, India

*Abstract* - Several schemes for gain control are used for preventing saturation of receiver, and overloading of data processor, tracker or display in pulse radars. The use of digital processing techniques open the door to a variety of digital automatic gain control schemes for analyzing digitized return signals and controlling receiver gain only at saturating clutter zones without affecting the detection at other zones. In this paper, we present a novel scheme of Digital Instantaneous Automatic Gain Control (DIAGC) which is based on storing digitally the dwell based clutter returns and deriving the gain control. The returns corresponding to first two PRTs in a dwell are used to analyze the presence of saturating clutter zones and the depth of saturation. Third PRT onwards proper gain control is applied at IF stage to prevent saturation of the following stages. FPGA based scheme is used for digital data processing, storing, threshold calculation and gain control generation. The effect of DIAGC on pulse compression is also addressed in this paper.

## I. INTRODUCTION

Clutter cancellation is a major problem in pulse radars with low-level coverage, especially when radar is required to detect low RCS target in a dense clutter environment. The finite dynamic range of any subsystem in the radar implies that the system will be saturated by large clutter returns i.e. the strength of clutter returns exceed the available dynamic range. In this condition, radar receivers with fixed gain exhibit objectionable characteristics. For MTI radars (using classical MTI canceller or Doppler filter bank), it has a severe effect on clutter cancellation in the signal processor. The un-cancelled clutter residues can result at the output of the delay line canceller producing large number of detections at the signal processor which eventually increases the load on data processor and floods the display screen.

The effect can be prevented by some well known techniques such as Sensitivity Time Control (STC), Automatic Gain Control (AGC), Instantaneous Automatic Gain Control (IAGC), raising the antenna beam etc. But all of them have their inherent limitations. Conventional clutter map formed apriory also may not hold good at the time of application (e.g. weather clutter).

In this paper, we present a novel scheme of DIAGC which can take care of surface clutter as well as volume clutter at required range, azimuth and elevation (r, θ, φ) within the coverage. The scheme prevents saturation of receiver only at the saturating clutter zones and doesn't affect the detection at other zones.

The remainder of the paper is organized as follows. Section 2 provides information about different gain control techniques including their limitations. Section 3 presents the DIAGC scheme including hardware implementation details and data processing operation. Section 4 shows subsystem and system measurements and plots. Section 5 shows the effect of DIAGC on pulse compression. Finally, section 6 provides several concluding comments followed by future plans in section 7.

## II. GAIN CONTROL TECHNIQUES

A radar must be designed to produce sufficient signal from the smallest target at its maximum range as well as to prevent saturation of receiver from large or close targets and clutters which in turn floods the display screen. When receiver saturates, it starts generating modulation products which spread into otherwise clutter free spectral region. This can be avoided by distributing the gain throughout the receiver chain with successive steps of gain control. From the point where gain control is applied, it helps prevent saturation in all following stages.

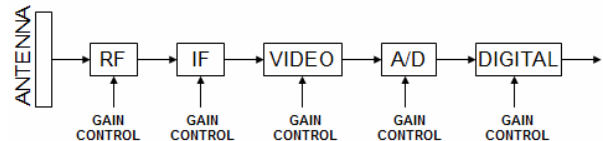

Fig. 1. Distribution of gain control

There are several techniques available to control the receiver gain, which are discussed along with their limitations in the following section.

### A. Sensitivity Time Control (STC)

STC is a technique by which the sensitivity of radar receiver is varied with time (range) to achieve more dynamic range. Maximum sensitivity is provided at long ranges, where it needs to detect the weak echoes of distant targets while the strong return from short ranges is prevented from saturating the system. Generally, this is achieved by applying attenuation in the RF front end. When the attenuation is applied at a rate 12 dB/octave in range, the amplified radar echo strength becomes independent of range. Other types of STC laws are also used in radar such as 9 dB/octave (or $R^3$ law) to handle surface clutter, 6 dB/octave (or $R^2$ law) to handle volume clutter etc.

The limitation of STC is that the medium PRF and high PRF pulse radars cannot employ STC to reduce clutter effects at short ranges unlike low PRF radars. The incompatibility of the STC requirements at elevation extremes severely limits the usefulness of STC especially in multi-beam radars. Moreover, STC cannot be applied at selective zones (r, θ, φ) where saturating clutter is present.

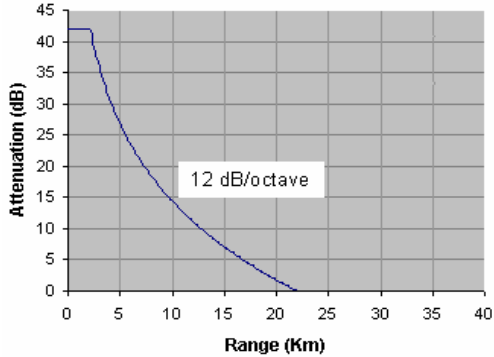

Fig. 2. Example of STC law.

*B. Automatic Gain Control (AGC)*

AGC is a technique which controls receiver gain automatically as and when the radar return signal changes in amplitude. The simplest type of AGC adjusts the receiver gain according to the average level of the received signal. Many surveillance radars employ an analog feedback loop configuration to achieve the required AGC action. AGC is generally applied at IF stages in a radar receiver.

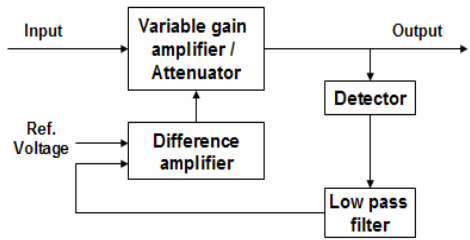

Fig. 3. Conventional AGC system.

The response time of conventional AGC systems limit its usefulness. If the received power varies rapidly, the AGC loop becomes unstable which can degrade tracking accuracy.

*C. Instantaneous AGC (IAGC)*

When AGC is performed on a pulse-to-pulse basis, it is called Instantaneous AGC (IAGC). AGC varies gain based on the return from a broad area while IAGC allows mapping of the high noise area. IAGC subtracts the power of the first pulse from the second; if the noise is uniform it will be erased from the receiver and display, leaving any targets that may be in the noise area.

III. DIAGC SCHEME

The DIAGC operates on dwell basis. A number of pulses are transmitted in a dwell towards a particular azimuth and elevation angle. The returns from each transmitted pulse are integrated over the dwell time and detection reports are generated in the signal processor. With the DIAGC, first two PRTs in a dwell are used to get an idea about the presence of saturating clutter returns and the depth of saturation. Third PRT onwards sufficient attenuation is applied (receiver gain is decreased) to prevent saturation in the following stages. On the other way, the DIAGC allows full amplification of weak signals and decreases the amplification of strong signals. As the attenuation controls are generated after digitizing the detected IF, hence the scheme can be called as DIAGC.

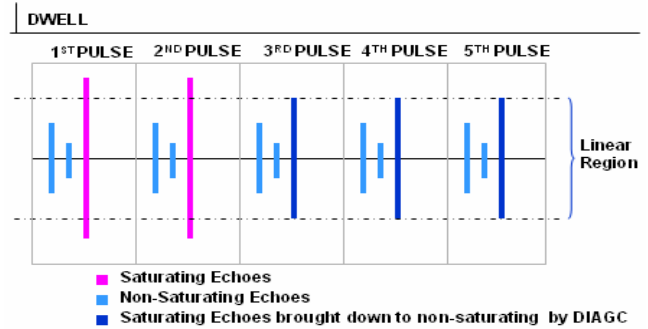

Fig. 4. DIAGC scheme

*A. RF Hardware Implementation*

Throughout the paper, we have considered a pulse radar with double down-conversion receiver. The receiver chain consists of RF front end, down converter and IF receiver. RF front end contains all waveguide components followed by low noise amplifiers (LNA). STC is applied at the RF front end before the LNA. The down converter section consists of mixers followed by IF amplifier prior to pulse compressors (Pre-IF). The IF receiver section consists of pulse compressors, post IF amplifiers and I/Q demodulators. The dynamic range available for gain control is maximum near the RF front end and decreasing towards I/Q detector. When only one IF stage is controlled, the range of DIAGC is limited to approximately 20 dB. When more than one IF stage is controlled, DIAGC range can be increased approximately by 40 dB.

For our study and implementation, we have selected the Pre-IF stage for DIAGC as the requirement for gain control was approximately 16 dB. The Pre-IF amplifier in the receiver is a gain controlled amplifier. The gain control is achieved by applying attenuation inside the pre-IF amplifier module. 6 bit control is available in the amplifier which corresponds to 31.5 dB of attenuation. These 6 bit controls are generated by DIAGC card, which is the digital hardware module for the DIAGC. The amplifier gain in the coupled path is adjusted in such a way that detector dynamic range matches with the gain control range.

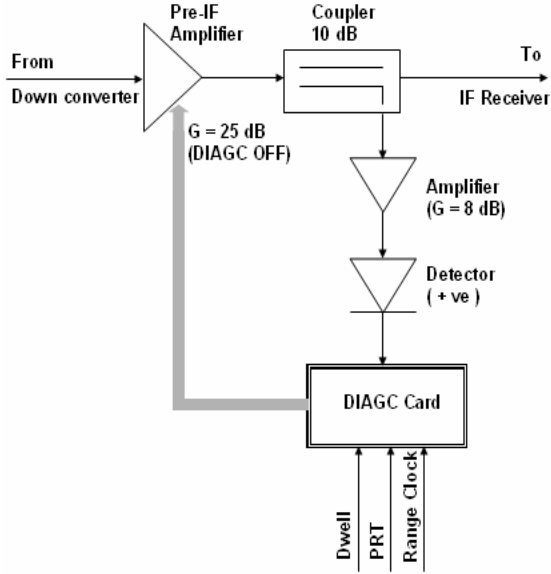

Fig. 5. RF hardware for DIAGC

*B. Digital Hardware Implementation*

The digital hardware for the DIAGC is a FPGA based card. It mainly contains a Xilinx Virtex FPGA (XCV200 PQ240) and a 14-bit ADC (AD 9240). This card receives dwell, PRT and range clock from the system timing generator through RS422 differential lines. The dwell and PRT keeps track of the number of PRT in a dwell. The range clock is the sampling clock for the ADC as well as the internal clock for the FPGA. Range-bin number is generated using the range clock. An onboard power on reset circuit provides the reset required for the FPGA. The FPGA is configured through an onboard serial PROM (XC18V04). The 14 bit sampled data from the ADC is sent to the FPGA for processing.

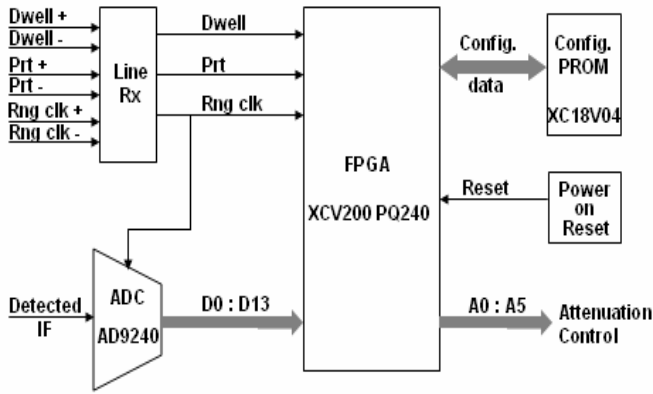

Fig. 6. DIAGC card block schematic

The 14 bit data received by the FPGA is passed through an 8 range-bin wide moving average filter for the first two PRTs in a dwell. During the first PRT, moving average data is stored inside the FPGA on range bin basis. During the next PRT, the stored data is read back and averaged with the current moving averaged data and again stored in the FPGA. Dual port block RAM cores are used as the memory element in the FPGA. Port A is used to write into the memory whereas port B is used to read from the memory. The control signals required for read and write operations are generated by the FPGA.

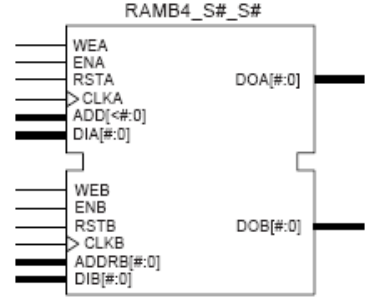

Fig. 7. Dual port block RAM core

When the third PRT comes, two PRT moving averaged data is read back from the memory and passed through threshold logic to generate proper attenuation control on range-bin basis. This process repeats for the rest of the PRTs in a dwell. Response time between the third PRT memory read operation and the change in Pre-IF amplifier gain is calibrated and correction factor is applied.

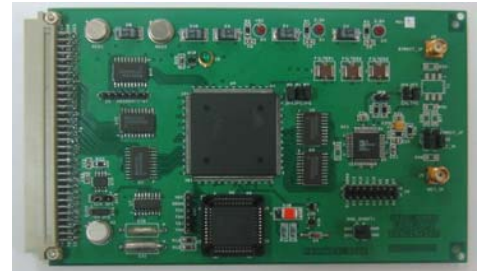

Fig. 8. DIAGC card

IV. RESULTS

The DIAGC operation was first tested in pulse radar with RF BITE (Built In Test Equipment). RF BITE is a simulated target injected at the RF front end for all azimuth and all elevation angles (although it can be restricted to any azimuth and elevation). For checking the DIAGC effect, the RF BITE was kept at a fixed range and the power level was adjusted in such a way that it saturated the receiver. When DIAGC was off, IF amplifier output and I/Q output waveforms were found distorted and the distortion was increasing with the increase in power level. As soon as the DIAGC was switched on, waveforms became distortion-less.

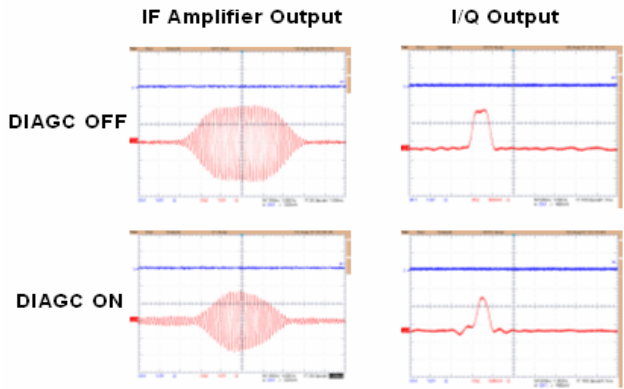

Fig. 9. DIAGC effect seen at different stages of receiver

The DIAGC effect was also seen at analog display with saturating RF BITE input. This test was carried out for different Doppler. When the Doppler was set to PRF/2 and DIAGC was off, a spread in range was observed around the specified range instead of single range-bin detection. As soon as the DIAGC was switched on, the spread vanished and single declaration was observed for each dwell. With zero Doppler, un-cancelled clutter residues due to saturation in the receiver were observed with DIAGC off. When the DIAGC was switched on, the display became clear. This experiment proves the display screen flooding due to saturation. And finally, no effect of DIAGC was observed when the system was in linear region.

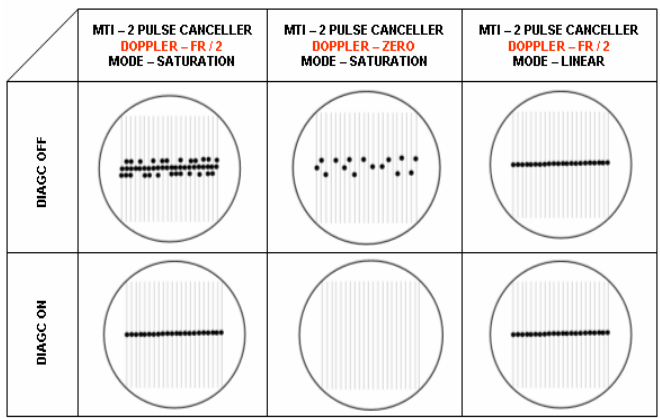

Fig. 10. DIAGC effect seen at analog display

Finally, the DIAGC effect was verified with actual radar transmission. When the radar was looking towards nearby hills, some of the hill returns was saturating the receiver with DIAGC off. The figure below shows the $2^{nd}$ and $3^{rd}$ PRT in zoomed condition. The captured clutter return shows both saturating and non-saturating clutters. When the DIAGC was off, both the $2^{nd}$ and $3^{rd}$ PRT consist of saturating clutter. When the DIAGC was switched on, the $3^{rd}$ PRT onwards there was no saturating clutter. $3^{rd}$ PRT onwards the receiver gain was adjusted to keep the receiver in the linear region.

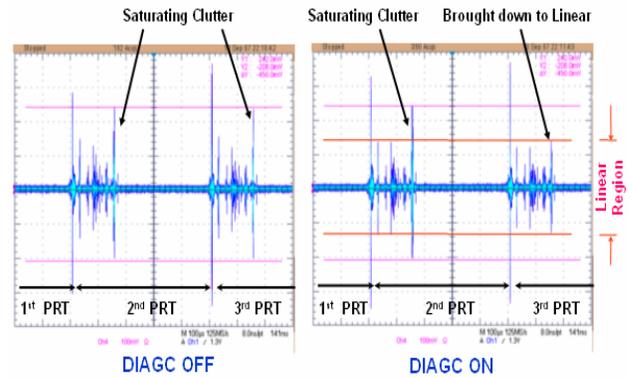

Fig. 11. DIAGC effect seen at receiver with transmission

Our DIAGC scheme can provide up to 16 dB of gain control which increases the dynamic range of the system by 16 dB. All the experiments were carried out in a working radar.

V. EFFECT OF DIAGC ON PULSE COMPRESSION

When step attenuation is applied on the uncompressed transmit coded waveform, the compressed pulse width and the time side-lobes increase which can affect the I/Q demodulation. The distortion on the compressed pulse width and the time side-lobes depend on the amount of attenuation applied in a step. For our case, maximum attenuation which can be generated is around 15 dB. For 8 range-bin moving average, the attenuation is applied at a rate of 2 dB/range-bin. The figure below shows some typical values of step attenuation applied on the uncompressed pulse and the distortion in the compressed pulse.

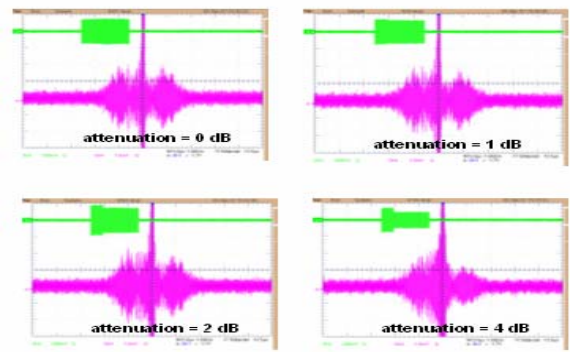

Fig. 12. Effect of step attenuation on pulse compression

VI. CONCLUSIONS

Based on results of this preliminary assessment, the use of DIAGC in any pulse radar is a promising approach for improving radar performance under saturating clutter environment. The key benefits of this approach are that it prevents saturation of receiver only at the saturating clutter zones without affecting the detection performance at other zones and this scheme can take care of surface clutter as well as time varying volume clutter. Moreover, the scheme is applicable for all high, medium and low PRF class of radars.

The main drawback of this approach is that it requires little more radar resources in terms of two extra PRTs in every dwell. One must be careful while choosing the attenuation step size prior to pulse compression.

## VII. FUTURE PLANS

The DIAGC scheme applied after pulse compression is a candidate approach for future experimental investigation that is currently being developed. Another approach comprising DIAGC at intermediate stages prior to IF in multiple conversion receiver shows promise for increased dynamic range, thereby improving radar system performance.